\documentstyle[eqsecnum,floats,preprint,epsf,aps]{revtex}
\tighten

\begin{document}

\renewcommand{\arraystretch}{1.5}
\newcommand{\be}{\begin{equation}}
\newcommand{\ee}{\end{equation}}
\newcommand{\bea}{\begin{eqnarray}}
\newcommand{\eea}{\end{eqnarray}}
\newcommand{\la}{\leftarrow}
\newcommand{\ra}{\rightarrow}
\newcommand{\lr}{leftrightarrow}
\newcommand{\La}{\Leftarrow}
\newcommand{\Ra}{\Rightarrow}
\newcommand{\Lr}{\Leftrightarrow}
\def\Tr{\mathop{\rm Tr}\nolimits}
\def\mapright#1{\smash{\mathop{\longrightarrow}\limits^{#1}}}
\def\mapdown#1{\big\downarrow \rlap{$\vcenter
  {\hbox{$\scriptstyle#1$}}$}}
\def\Z{{\bf Z}}
\def\R{{\bf R}}
\def\M{{\cal M}}
\def\L{{\cal L}}
\def\c{c_\chi}
\def\s{s_\chi}
\def\xh{\hat x}
\def\yh{\hat y}
\def\zh{\hat z}
\def\vt{\widetilde{v}}
\def\wt{\widetilde{w}}
\def\n#1{{\hat{n}_{#1}}}
\def\c#1{\cos{#1}}
\def\s#1{\sin{#1}}
\def\cs#1{\cos^2{#1}}
\def\ss#1{\sin^2{#1}}
\newcommand{\PSbox}[3]{\mbox{\rule{0in}{#3}\includegraphics{#1}\hspace{#2}}}

\title{Stacking non-BPS D-Branes}

\author{Gian Luigi Alberghi $^{1}$,  Elena C\'aceres$^{2}$, Kevin Goldstein$^{3}$ and David A. Lowe$^{3}$}

\address{~\\$^1$Dipartimento di Fisica, Universita' di Bologna and \\
I.N.F.N, Sezione di Bologna \\
Via Irnerio,46, 40126 Bologna, Italy. \\
{\tt alberghi@bo.infn.it}}

\address{~\\$^2$Abdus Salam ICTP\\
Strada Costiera 11, 34014 Trieste, Italy. \\
{\tt caceres@ictp.trieste.it}}

\address{~\\$^3$
Brown University \\
Providence, RI 02912, USA. \\
{\tt kevin@het.brown.edu}, 
{\tt lowe@het.brown.edu}}

\maketitle

\begin{abstract}

We present a candidate supergravity solution for a stacked configuration of stable 
non-BPS D-branes in Type II string theory compactified on $T^4/Z_2$. 
This gives a supergravity description of nonabelian tachyon 
condensation on the brane worldvolume.

\end{abstract}

\setcounter{page}{0}
\thispagestyle{empty}

\vfill

\noindent BROWN-HET-1263\hfill\\
\noindent IC/2001/27 \hfill\\

\eject

\vfill

\eject

\baselineskip 20pt plus 2pt minus 2pt

\section{Introduction}
\label{sec:introduction}

The seminal work of Sen \cite{Sen1,Sen2,Sen3}  underlined the 
importance of understanding 
stable  non-BPS branes in string 
theory. Non-BPS branes provide us with new ways to construct stable
non-supersymmetric states in string theory, which ultimately may lead
to realistic brane world models as exact solutions. 
It seems likely they will lead to new insights into the outstanding
phenomenological difficulties of string theory. Review articles on
non-BPS branes may be found in \cite{reviews}.

In this paper we consider a class of non-BPS branes with an integer
valued conserved charge \cite{Sen:1998ex,Bergman:1999kq} that arise
from non-BPS branes on a $T^4/Z_2$ orbifold of Type II string
theory. The $Z_2$ symmetry acts by reflection in the four toroidal
directions combined with $(-1)^{F_L}$ ($F_L$ is the left-moving
fermion 
number).
The orbifolding removes the open string tachyons, making the
configuration  stable. The brane is charged with respect to a
twisted sector Ramond-Ramond gauge field of the orbifolded Type II
string theory. States that only carry charge under this gauge field
will not be supersymmetric, as there is no associated central charge.

If the radii of the torus are fixed to a particular value the branes
do not interact 
at first order in 
perturbation theory and there is a no-force condition similar to
that of the BPS case \cite{gabsen}. 
One expects there to exist a 
classical supergravity solution when a 
large number of such branes are stacked on top
of each other. Studies of this supergravity solution appear in
\cite{DiVecchia,Eyras,Lozano,Bain}. Related supergravity solutions for
coinciding branes and anti-branes have appeared in \cite{doublezu,Brax}. In particular,
\cite{DiVecchia,Bain} construct supergravity solutions with
the appropriate gauge charge, but with naked singularities.
These solutions
violate the scalar no-hair theorems of general relativity. 
One therefore expects such solutions will be unstable. This was
confirmed in \cite{DiVecchia}, where a repulsive force was found to
act on a probe non-BPS brane in their classical background solution.

Starting with the six-dimensional supergravity action arising
from Type II string theory on $K3$, we find solutions compatible with
the no-hair theorems with the same conserved charge as the
configuration of non-BPS branes. These are argued to be the
classical solution corresponding to the
stable ground state of the coinciding non-BPS branes. 
The solutions are six-dimensional analogues of the 
well known black p-branes (see
for example \cite{DuffandLu,Stelle,HoroStro,GibbMaed}). 
The solutions we construct are free of naked
singularities.

In \cite{Lambert} the worldvolume theory of coinciding stable non-BPS
branes has been studied. Although individual branes are tachyon-free,
non-abelian tachyons appear when branes are coincident. It was argued
there should be a stable ground state for such a configuration of
branes, where these tachyons condense at a nontrivial minimum. The
supergravity solution we construct will presumably correspond to this
stable minimum. 
The tension to charge ratio we find is a factor of $\sqrt{2}$
larger than one would expect starting from the perturbative string
theory result \cite{Sen:1998ex,Bergman:1999kq}. This yields a
nontrivial prediction for the strongly coupled groundstate of the
worldvolume theory of the non-BPS branes.

\section{General Black Brane Solutions}
\label{sec:genact}
In this section we review the construction of general black brane
solutions in arbitrary dimensions (see, for example, \cite{doublezu}).
The starting point for the description of a brane charged
with respect to one gauge field in a D-dimensional spacetime is
the action 
\be
\label{generalaction}
S = \int d^D x \sqrt{-g} \left[ R - {1 \over 2} \nabla^M \phi
  \nabla _M \phi  - 
  {1 \over 2(p+2)!} e^{a \phi} F_{[p+2]} ^2 \right]~,
\ee 
where $x^M, M=0,..D-1$ are the spacetime coordinates, 
$g$ is the determinant of the metric, $R$ the Ricci scalar,
$\phi $ is the dilaton field, $F_{[p+2]}$ is the $(p+2)$-form
field strength of the $(p+1)$-form gauge potential $A_{[p+1]}$, 
that is $ F_{[p+2]}= d A_{[p+1]} $ and $a$ is a constant governing
the coupling between the dilaton and the gauge potential.
In general the effective action contains the metric tensor,
various ranks of antisymmetric tensor fields and many scalars.
However it can be shown that
(\ref{generalaction})
is a consistent truncation of the whole action, so that 
the solutions of the equations of motion obtained by varying this
action are particular solutions of the full theory.
The equations of motion obtained by varying the above action
can be expressed as
\bea
\label{generaleqofmotion} 
R_{MN} &=& {1 \over 2} \nabla _M \phi \nabla _N \phi + S_{MN}
\nonumber \\
\nabla _{M}  \left( e^{a \phi} F^{MM_1..M_{p+1} } \right) &=& 0
\nonumber \\
\Box \phi &=& { a \over 2 (p+2)!} e^{a \phi} F^2  ~,            
\eea
with
\be
S_{MN} = { 1 \over 2(p+1)! } e^{a \phi} \left(
         F_{M M_1... M_{p+1}} F_N ^{M_1...M_{p+1} } -
         { p+1  \over (p+2)(D-2) } F^2 g_{MN} \right)~.
\ee               \\
Looking for the general black p-brane solutions of these equations we do not require 
Poincar\'e invariance on the $d\equiv p+1$ world-volume of the brane.
We look instead for the most general solutions that preserve the symmetry
\be
\label{symmetry}
   {\cal{S}} = SO(p) \times SO(D-p) ~,
\ee
that is a rotational symmetry in the spatial directions transverse to the
brane and on the brane.
We can split the spacetime coordinates as $X^M= (x^{\mu},y^m)$
where $x^{\mu}$ $ (\mu=0,..,d-1) $ are the spatial coordinates on the brane
worldvolume and $y^m$ $(m=d,..,D-1) $ are the coordinates on the
transverse spacetime directions. 
The ansatz for the metric, given the symmetry (\ref{symmetry}), is
\be
\label{genmetricansatz}
  ds^2 = e^{2A(\rho)} \left( -f(\rho) dt^2 + \eta _{\alpha \beta} 
         dx^{\alpha} dx^{\beta}  \right)  +
         e^{2B(\rho)} \left( \delta_{mn} dy^m dy^n \right)~,
\ee
where $\alpha =1,...,d-1$ and $ \rho \equiv \sqrt{y^m y^m} $ is 
the isotropic radial coordinate
in the transverse space. The ansatz for the dilaton is simply
$ \phi(x^M) = \phi(\rho) $ and that for the gauge potential, for the case
of an electric brane, is
\be
\label{generalgaugeansatz}
    A _{\mu_1 \mu_2 ... \mu_{d} }=  e^{\Lambda(\rho)} 
    \epsilon _
{\mu_1 \mu_2 ... \mu_{d} } \;\; 
    \mbox{, other components zero,}
\ee
where  $\epsilon _
{\mu_1 \mu_2 ... \mu_ {d}}$ is the usual 
antisymmetric tensor ($\epsilon _
{0 1 ... d } =1$).
The equivalent ansatz in terms of the field strength is
\be
    F_{m \mu_1 \mu_2 ... \mu_{d}} =  
    \epsilon _
{\mu_1 \mu_2 ... \mu_{d} } \partial _m
       e^{\Lambda (\rho)}  \; \; \mbox{, other components zero.}
\ee
The general solution with this ansatz is constructed in
\cite{doublezu}. 
If we require an asymptotically flat metric and a regular horizon,
we find the solutions: 
\bea
\label{blackbrane} 
ds^2 & =& - \left[ 1- \left( { r_{+}\over r} \right)^{\tilde d} \right]
         \left[ 1- \left( { r_{-}\over r} \right)^{\tilde d} \right]^
         { {4 \tilde d \over \Delta(D-2)} -1} dt^2                         
+         \left[ 1- \left( { r_{-}\over r} \right)^{\tilde d} \right]^
         { {4 \tilde d \over \Delta (D-2)}} dx^{\alpha} dx_{\alpha}
      \nonumber   \\
& & \qquad
+  \left[ 1- \left( { r_{+}\over r} \right)^{\tilde d} \right] ^{-1}
   \left[ 1- \left( { r_{-}\over r} \right)^{\tilde d} \right] ^
        { {2 a^2 \over \Delta \tilde d} -1} 
        d r ^2         
+ r ^2 
\left[ 1- \left( { r_{-}\over r} \right)^{\tilde d}
 \right] ^
{{2 a^2 \over \Delta \tilde d}} d \Omega ^2 _{\tilde d+1}     \nonumber        \\
e^ {\phi} &=& \left[ 1- \left( { r_{-}\over r} \right)^{\tilde d}
         \right] ^ {-{2a \over \Delta}}
         \nonumber \\
F &=& \lambda \ast \epsilon _{D-p-2}~,
\eea
where $r_\pm$ are the positions of the inner and outer horizons, 
$\epsilon_{n}$ is the volume form on the unit $n$ sphere,
$\tilde d=D-p-3$,
\be
\Delta = {2 d \tilde d \over D-2} +a^2~,
\ee
and the charge $\lambda$ is
\be
 \lambda = {2 \tilde d \over \sqrt{\Delta}} (r_{-} r_{+})^{\tilde d \over 2}~.
\ee
In writing this solution we have changed to Schwarzschild coordinates
via $\rho^{\tilde d}= r^{\tilde d}-r_-^{\tilde d}$.
In components the field strength is
\be
F_{m \mu_1 .. . \mu_{n-1} }= -{2 \over \sqrt{\Delta} } 
 \epsilon_{\mu_1 ... \mu_{n-1} } \partial _m 
\left( {r_{-} \over r } \right) ^{\tilde d} ~.
\ee
In the extremal limit $r_+=r_-$.
These are the expressions for the metric tensor, the dilaton
and the field strength for a black p-brane in a D-dimensional
spacetime. We will show that these are the relevant solutions for
the problem we are discussing.

\section{Non-BPS brane solutions of Type II supergravity on $T^4/Z_2$}
\label{sec:aefe}

The starting point for the analysis of the particular case we 
are addressing, 
is the action coming from a truncation of the ten-dimensional
type II supergravity theory compactified on an orbifolded 
torus $T^4 / Z_2 $. We will be interested in black p-brane solutions
related to non-BPS branes. We therefore take $p$ even for Type IIB,
and $p$ odd for Type IIA.
Actually we will examine a truncation of the full
six-dimensional gravity theory describing the metric, the gauge potential
and five scalars as discussed in \cite{DiVecchia}.
The truncated action in Einstein frame is 
\be
\label{action}
S = \int d^6 x \sqrt{-g} \left[ R - {1 \over 2} \left(\nabla \phi
  \right) ^2 - {1 \over 2} \left(\nabla \eta_b \right) ^2 -
  {1 \over 2 (p+2)!} e^{a \phi} F ^2_{[p+2]} \right]~,
\ee  
where $g$ is the determinant of the six dimensional metric,
$\phi$ is the dilaton field,$ \eta _b$ with $b=1,..,4$ are scalar fields 
associated with the size of the $T^4$ and
$ F _{[p+2]}$ is the $(p+2)$-form field strength for a $(p+1)$-form gauge potential 
$A_{[p+1]}$. \footnote{In order to recover  the conventions of
  appendix B of \cite{DiVecchia} we must renormalize $\varphi = \phi/
  \sqrt{2}$, along with a compensating rescaling of $a$.}
Here $a=(1-p)/\sqrt{2}$.

Varying this action one obtains equations of motion
with the structure
of (\ref{generaleqofmotion}) with additional terms involving
derivatives of the scalars $ \eta_b$, and the additional equations of motion
\be
\Box \eta_b  =0~.
\ee

First, we note that   
the scalars $ \eta _b $ have an equation of motion
that can be integrated once immediately.
No hair theorems show that the only scalar field which approaches a constant
at infinity and is regular at the event horizon of a black brane
solution is everywhere constant \cite{nohair}.  Otherwise the solution
will have a naked singularity, and in line with the cosmic censorship
conjecture, it is expected 
such a solution will generically be
unstable.
As we are looking for a non-singular solution, or at least a
solution with singularities hidden by a horizon, we can 
set the $\eta _b $ to constants. They then decouple from the other
equations of motion.
This is an important step toward the solution of the problem, 
since we are now left with the equations of motion examined in the
first section, that leads to the black brane 
solutions for a six-dimensional spacetime. Note that the $\eta_b$ are
non-constant for the elementary non-BPS brane \cite{Eyras}. 
We therefore propose that once the nonabelian open-string tachyons on
the worldvolume of the branes have condensed to a stable minimum, 
the $\eta_b$'s decouple from the solution. The critical radius of the
elementary non-BPS brane, below which it becomes unstable, corresponds
to $\eta_b=0$.

We can carry over the results of the previous section, identifying
$D=6$ and $\Delta=2$, which leads to $\tilde d=3-p$. We will focus on
the cases $p<3$.
Using the results of the previous section,
we can straightforwardly write down the solutions
\bea 
\label{nonextremal}
ds^2 &=& - \left[ 1- \left( { r_{+}\over r} \right)^{3-p} \right]
         \left[ 1- \left( { r_{-}\over r} \right)^{3-p} \right]
         ^{(1-p)/2}
dt^2    
+ \left[ 1- \left( { r_{-}\over r} \right)^{3-p} \right]
         ^{(3-p)/2} dx^{\alpha} dx_{\alpha} \nonumber \\
& & \qquad + \left[ 1- \left( { r_{+}\over r} \right)^{3-p} \right] ^{-1}
   \left[ 1- \left( { r_{-}\over r} \right)^{3-p} \right] ^
        { p^2-5 \over 2(3-p)} d r ^2          
+ r ^2 \left[ 1- \left( { r_{-}\over r} \right)^{3-p}
         \right] ^
{(1-p)^2 \over 2(3-p)} d \Omega ^2 _{5-d}                  \nonumber    \\
e^ {\phi} &=& \left[ 1- \left( { r_{-}\over r} \right)^3 \right]
         ^ {(p-1)\over \sqrt{2}}                             \nonumber  \\
F &=& \lambda \ast \epsilon _{5-d}~,
\eea
and $\lambda = \sqrt{2} (3-p) (r_+ r_-)^{(3-p)/2}$.

This solution depends on only two parameters $ r_{+}$ and $r_{-}$ which are
related to the mass and gauge charge,
consistent with the no hair theorems. We can explicitly see that 
the singularity located at
$r_{-}$ is hidden by an horizon at $r_{+}$.
At this stage this is a black brane solution of the
six-dimensional theory.  The Hawking
temperature of these solutions is
\be
\label{temperature}
   T = {3-p \over 4 \pi r_{+} }  
   \left [ 1- \left( { r_{-} \over r_{+} } \right) ^{3-p} \right]
   ^ {{p-2 \over p-3}} ~.
\ee
This leads to a positive specific heat in the extremal limit for
$p=0,1$, and a negative specific heat for $p=2$. We therefore expect
the black membrane will  suffer from a Gregory-Laflamme 
instability \cite{Gregory:1994bj}, but the
black D-particle ($p=0$) solution and the black D-string should be stable. 

We can take the extremal limit of the solution (\ref{nonextremal}),
by setting $r_{+} = r_{-} \equiv {r_0} $
\bea 
\label{extremal}
ds^2& =& - 
\left[ 1- \left( { r_{0}\over r} \right)^{3-p} \right]
         ^{3-p \over 2}
\left(dt^2    
+  dx^{\alpha} dx_{\alpha} 
   \right)   \nonumber   \\                  
& & \qquad + \left[ 1- \left( { r_{0}\over r} \right)^{3-p}
         \right] ^
{p^2+2p-11 \over 2(3-p)} 
\left( d r ^2   + 
\left[ 1- \left( { r_{0}\over r} \right)^{3-p}
         \right] ^
{2}  r ^2 
d \Omega ^2 _{5-d}         \right)         \nonumber    \\
e^ {\phi} &=& \left[ 1- \left( { r_{0}\over r} \right)^{3-p}\right]
         ^ {p-1\over \sqrt{2}}                              \nonumber \\
F& =& \lambda \ast \epsilon _{5-d}~.
\eea
The extremal solution is Poincar\'e invariant along the D-brane
worldvolume as expected. The unstable modes of \cite{Gregory:1994bj} become
degenerate in the extremal limit, and entropy arguments suggest the extremal 
state should be stable \cite{Gregory:1995tw}. 
This solution will be our candidate for the stable
end-point of tachyon condensation on the worldvolume of coincident
non-BPS D-branes. The solution  has a singular horizon in the extremal
limit (rather than a naked singularity), 
which is typical of dilatonic BPS solutions.
Taking the extremal limit in the expression (\ref{temperature}) 
for the cases $ p=0,1 $ one can see that the temperature goes to zero, 
whereas in the case $ p=2 $ it tends to a the finite value $1 / 4 \pi r_{0}$.
\par 
We can compute the mass and charge of the solution by examining the
asymptotic behavior of the fields at infinity:
\bea
g_{00} &\sim& -\left( 1- {3-p\over 2} \left({ r_0 \over r}\right)^{3-p}
\right) \nonumber \\
A_{0\cdots p} &\sim& \sqrt{2}\left( 1- \left({ r_0 \over r}\right)^{3-p}\right)~.
\eea
If we demand the charge matches that of $N$ coincident branes, then
comparing to \cite{DiVecchia}, we find a tension  larger by a factor $\sqrt{2}$
than that of an elementary non-BPS brane. Recall the elementary
non-BPS brane has the same tension as a BPS D-brane, but two units of
charge \cite{Eyras}. 
We interpret the change in
the tension of the stacked branes 
as a result of tachyon condensation on the worldvolume. Because the
solution is non-supersymmetric, we also expect renormalization of the
tension  versus the perturbative result.

The Born-Infeld action for a static D p-brane probe in this background takes
the form
\be
S =-M \int d^{p+1} \xi \   e^{-{a\phi /2} - \sum_b
  \eta_b/2 } {\sqrt{ -{\rm det}
g_{\alpha \beta}}} + Q\int A_{p+1}~.
\ee
Where $M$ is the probe tension, and $Q$ is proportional to the probe charge.
Inserting  the extremal background solution, and setting the
$\eta_b=0$ corresponding to the critical radius of the $T^4$
directions for convenience, one obtains
\be
\label{biact}
S= -\int d^{p+1} \xi \left(1- \left({r_0\over r}\right)^{3-p}\right)
\left(M -{\sqrt{2} Q}\right)~.
\ee
If we take the probe to be a set of stacked branes, 
$M/Q=\sqrt 2$, the static potential vanishes, as one might expect for
an extremal black hole solution. However, if we take $M/Q=1$ as
appropriate for an elementary non-BPS probe brane, we find a
repulsive force. We can therefore expect the stacked configuration to
slowly discharge due to pair production of elementary non-BPS branes
via the Schwinger mechanism. As is apparent from (\ref{biact}) the
force on the pair will be very small near the horizon of a large black
hole.  The  rate of production will go like $\exp(-{\rm const.} ~M^2
r_0)$.
The black hole will be
 absolutely stable only
in the limit of a
large number of branes.

\section{Conclusions}

Starting from an action describing a consistent truncation of 
ten dimensional type II superstring theory compactified on an orbifolded 
four dimensional torus $T^4/Z_2$ we have considered 
stable classical solutions describing a stack of non-BPS
D-branes. We have argued the stable solution
should correspond to the simplest extremal black hole solution with
the same gauge charge and spacetime symmetry as the set of
branes. This yields a nontrivial prediction for the stable groundstate
of the nonabelian tachyon condensation on the worldvolume of the
stacked branes. It seems likely these ideas may be generalized
straightforwardly to other brane systems where one expects a
supergravity description of tachyon
condensation to a stable ground state \cite{Brax}. 

\section*{Acknowledgments}
G.L.A. would like to thank Brown HEP theory group for hospitality
and in particular Steven Corley for helpful discussions.The work of
E.C. was supported in part by the European Commission  grant
ERBFMRX-CT96-0090. The research of K.G. and D.L. is supported in part by DOE
grant DE-FE0291ER40688-Task A. D.L. thanks R. Myers for helpful discussions.

\end{document}